\documentclass{ws-procs9x6}

\newcommand{\half}{\frac{1}{2}}
\newcommand{\fm}{\textrm{fm}}
\newcommand{\gev}{\textrm{GeV}}
\newcommand{\qs}{Q_\mathrm{s}}
\newcommand{\ls}{\Lambda_\mathrm{s}}
\newcommand{\lqcd}{\Lambda_\textrm{QCD}}
\newcommand{\pp}{p_\mathrm{T}}
\newcommand{\kk}{k_\mathrm{T}}
\newcommand{\ca}{C_\mathrm{A}}

\newcommand{\ra}{R_A}
\newcommand{\gt}{\gamma^0}
\newcommand{\gz}{\gamma^3}
\newcommand{\dtau}{\partial_\tau}
\newcommand{\ud}{\textrm{d}}
\newcommand{\cf}{C_\mathrm{F}}
\newcommand{\y}{y_\mathrm{T}}
\newcommand{\xx}{x_\mathrm{T}}
\newcommand{\nr}[1]{(\ref{#1})}

\begin{document}

\title{Classical fields and heavy ion collisions}

\author{T. Lappi}

\address{Department of Physical Sciences,
Theoretical Physics Division
 and \\
Helsinki Institute of Physics \\
P.O. Box 64,
FIN-00014 University of Helsinki, FINLAND\\
E-mail: tuomas.lappi@helsinki.fi}

\maketitle

\abstracts{
This is a review of numerical applications of classical gluodynamics
to heavy ion collisions.
We recall some results from calculations of gluon production, discuss their
implications for heavy ion phenomenology, and outline
a strategy to calculate the number of quark pairs produced by these classical
fields.}

\section{Introduction}

To understand what is happening in relativistic heavy ion colliders,
currently RHIC in Brookhaven and  in the future the LHC at CERN, one needs
to understand not only hard probes, but also bulk particle production and
thermalisation.
It can be argued that the large phase space densities of partons in the
small-$x$ wavefunction of the nuclei generate a hard enough momentum scale
to allow weak coupling methods to be used.
We shall discuss first some general ideas concerning weak coupling, classical field
methods in this context. Then we will go on to discuss first gluon and then
quark pair production in the McLerran-Venugopalan model.

\section{Relativistic heavy ion collisions}

We shall be interested in studying the case where two nuclei move at the
speed of light along the $x^\pm=0$-axes, i.e. at $\tau=0$\footnote{
The light cone coordinates are defined as $ x^\pm = (t \pm z)/\sqrt{2} $
 and the proper time and spacetime
rapidity as: $\tau = \sqrt{t^2-z^2}$ and $\eta = \half \ln \frac{x^+}{x^-}$.
}.
These nuclei then collide and leave behind them, at finite
values of $\eta$ and $\tau$, some matter which is then observed in
detectors located in some region, varying between different experiments,
around $\eta=0$.
We can divide the collision process in different stages:
\begin{enumerate}
\item The initial condition at $\tau = 0$ depends on the properties
of the nuclear wavefunction at small $x$. 
\item  The thermal and chemical equilibration of the matter formed
at $\tau \lesssim \tau_0$ requires understanding
of time dependent, nonequilibrium Quantum Field Theory.
\item The Quark Gluon Plasma, surviving for some fermis
around $\tau_0 \lesssim\tau \lesssim 10 \fm$. 
If the system reaches local thermal equilibrium, finite temperature 
field theory and relativistic hydrodynamics
can be used to describe its behaviour.
\item Finally, for $\tau  \gtrsim 10 \fm$ the system hadronises and decouples.
\end{enumerate}

The question of the thermalisation timescale $\tau_0$ remains poorly understood.
Hydrodynamical calculations have bees very successful in explaining
the experimental observations, but their success depends on the assumption
of a very early thermalisation time\cite{kolb}.
Most perturbative estimates, i.e. the bottom-up scenario\cite{Baier:2000sb},
generically produce quite a large thermalisation time, $\tau_0 \gtrsim 3 \fm$.
On the other hand, one could argue that if the behaviour of the 
system is characterised by some quite large momentum scale, i.e. the saturation
scale $\qs \sim 1 \ldots 2 \gev$, thermalisation should occur already at times
$\tau_0 \sim 1/\qs \sim 0.2 \fm$. It has been pointed out recently 
(see e.g. Ref.~\refcite{arnold}) that plasma instabilities could provide the rapid
thermalisation that hydrodynamical models require.
In this context classical field models of the nuclear wavefunction and
particle production can provide some insight into understanding
the collision process.

\section{Saturation and the classical field model}

The general idea of parton saturation is that the the small-$x$ components 
of the nuclear wavefunction are dominated by a transverse momentum scale,
the \emph{saturation scale}
$\qs$. The scale $\qs$ is supposed to grow for decreasing $x$
as $\qs^2(x) \sim x^{-\lambda}$ with $\lambda \approx 0.3$ giving a
good fit to HERA data on deep inelastic scattering\cite{Golec-Biernat:1998js}.
Thus for small enough $x$ or, equivalently, large enough $\sqrt{s}$
we have $\qs \gg \lqcd$ and we can use weak coupling methods.
The presence of the scale $\qs$ is supposed to be caused by the color fields
in the nuclear wave function becoming so strong that the
gluonic interactions are dominated by the nonlinearities in the
Yang-Mills Lagrangian.

This idea can also be thought of as parton percolation. Let us
attach, arguing by the Heisenberg uncertainty principle, 
 to an individual parton with transverse momentum $\pp$
a transverse area $\sim 1/\pp^2$. One expects a qualitative change in the
behaviour of the system at momentum scales where the partons 
overlap and percolate the transverse plane:
$N / \pp^2 \sim \pi \ra^2 $, where $N$ is the number of partons and
$\pi \ra^2$ the nuclear transverse area.

Because the color fields are strong, the occupation numbers of quantum
states of the system are large, and it is natural ot use a classical field 
approximation.
The saturation model has been dubbed  the ``Color Glass Condensate'' and
even called ``a new form of matter''.
In theory the term ``color glass condensate'' refers to the small-$x$ wavefunction
of the nucleus, characterised by the saturation scale $\qs$. In
practice, most applications of these ideas to heavy ion collisions
have so far not been classical field computations, but 
perturbative calculations
with some phenomenological  gluon $k_\mathrm{T}$-distribution
depending on $\qs^2(x).$

\subsection{Heavy ion collisions in the classical field model}

To study particle production in the classical field model we have to 
solve the Yang-Mills equations of motion:
\begin{equation}\label{eq:ym}
[D_{\mu},F^{\mu \nu}] = J^{\nu},
\end{equation}
with a current of two infinitely Lorentz-contracted nuclei moving
along the two   light cones:
\begin{equation}\label{eq:source}
J^{\mu} =\delta^{\mu +}\rho_{(1)}(\xx)\delta(x^-) 
+ \delta^{\mu -}\rho_{(2)}(\xx)\delta(x^+).
\end{equation}
The model part of this approach comes when one must take some form
for the charge density $\rho(\xx)$. The suggestion of McLerran and
Venugopalan\cite{McLerran:1994ni} 
was to take a random stochastic color source with a Gaussian distribution:
\begin{equation}\label{eq:korr}
\langle \rho^a(\xx) \rho^b(\y) \rangle = g^2 \mu^2 \delta^{ab}\delta^2(\xx-\y)
\end{equation}
and then average all quantities calculated from $\rho(\xx)$ 
with this distribution. An important development has been 
to consider a more general $x$-dependent 
probability distribution $W_x[\rho(\xx)]$ and derive a 
renormalisation group equation for this distribution
as a function of $x$, the JIMWLK equation\cite{weigert}.

\subsection{Note on different saturation scales}

Different ways of understanding saturation have led to different definitions
of the saturation scale in the literature, which can be a source of 
considerable confusion.

\begin{itemize}
\item The $p_\mathrm{sat}$ in the EKRT\cite{Eskola:1999fc} model is a 
final state saturation scale, not a property of the nuclear wave function
(initial state), and thus conceptually a bit different from
the saturation scale we have discussed so far.
\item We are using the old notation of of Krasnitz et. al.
In their newer work\cite{Krasnitz:2002mn} they define
 $\ls = g^2 \mu$.
\item By calculating the gluon distribution in the McLerran-Venugopalan
model one can relate the strangth of the color source to the saturation
scale $\qs$ defined by A. Mueller  and Yu. Kovchegov
(see e.g. Refs.~\refcite{Krasnitz:2002mn,Kovchegov:2000hz})\footnote{
Note the dependence on an infrared cutoff $\lqcd$ that gives
a large numerical uncertainty.}:
\begin{equation}\label{eq:satscale}
\qs^2 = \frac{g^4 \mu^2 \ca }{4 \pi} 
\ln \left( \frac{g^4 \mu^2}{\lqcd^2} \right) 
\end{equation}
\item The saturation scale, or radius, $\qs = 1/R_\mathrm{s}$ in the work of
Golec-Biernat and Wusthoff\cite{Golec-Biernat:1998js}
or Rummukainen and Weigert\cite{Rummukainen:2003ns} is the same 
except with $\ca$ replaced by $\cf$.\footnote{
The techical reason for this is that they consider correlators
of Wilson lines in the fundamental representation, whereas the the
gluon distribution that Mueller and Kovchegov consider involves
the same correlator in the adjoint representation.}
\item For a comment on the relation to the saturation scale used by 
E. Iancu et. al., see Ref.~\refcite{Lam:2002mg}.
\end{itemize}

\section{Gluon production in the classical field model}

\begin{minipage}[b]{0.39\textwidth}
The solution of the Yang-Mills equations in regions
(1) and (2) is an analytically known pure gauge field\cite{Kovner:1995ts},
and gives the initial condition for the numerical solution in the
forward light cone (3).
\end{minipage}
\includegraphics[width=0.59\textwidth]{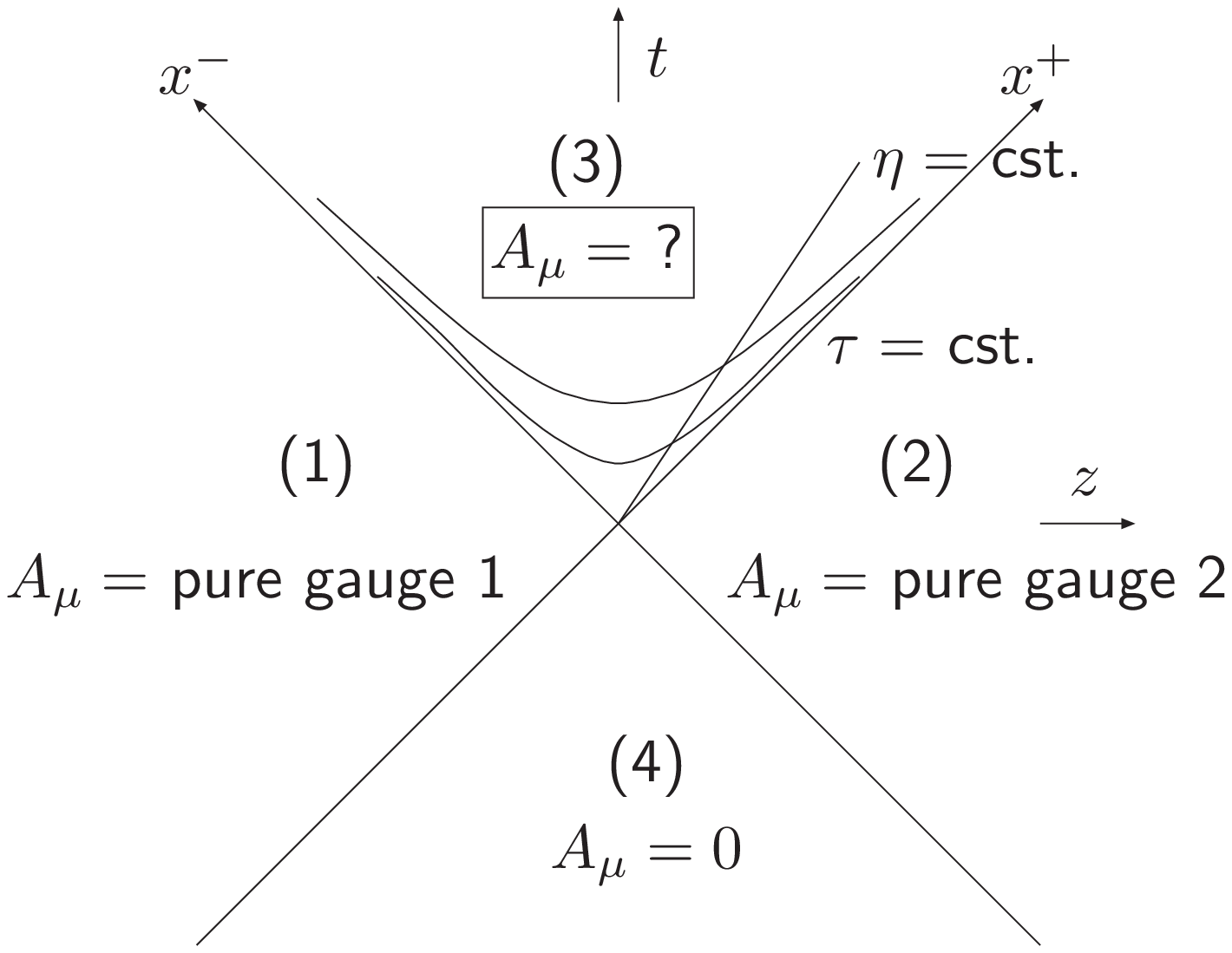}
The numerical method for solving the Yang-Mills equations in the forward
light cone was developed by Krasnitz, Nara and
Venugopalan\cite{Krasnitz:1998ns,Krasnitz:2001qu}.
The correct result for the transverse enegy was found in Ref.~\refcite{Lappi:2003bi},
(see also the erratum to Ref.~\refcite{Krasnitz:2001qu} in Ref.~\refcite{Krasnitz:2003jw}).

The numerical computation is done in the Hamiltonian formalism. Due to the
boost invariance of the initial conditions the Yang-Mills equations
can be dimensionally reduced to a 2+1 dimensional gauge theory with an adjoint
scalar field. With the assumption of boost invariance one is explicitly
neglecting the longitudinal momenta of the gluons; a restriction
that should be relaxed in future computations. In the Hamiltonian
formalism one obtains directly the (transverse) energy. By decomposing
the fields in Fourier modes one can also define a gluon multiplicity
corresponding to the classical gauge fields.

\subsection{Numerical results}

\begin{figure}
\begin{center}
\noindent
\includegraphics[width=0.49\textwidth]{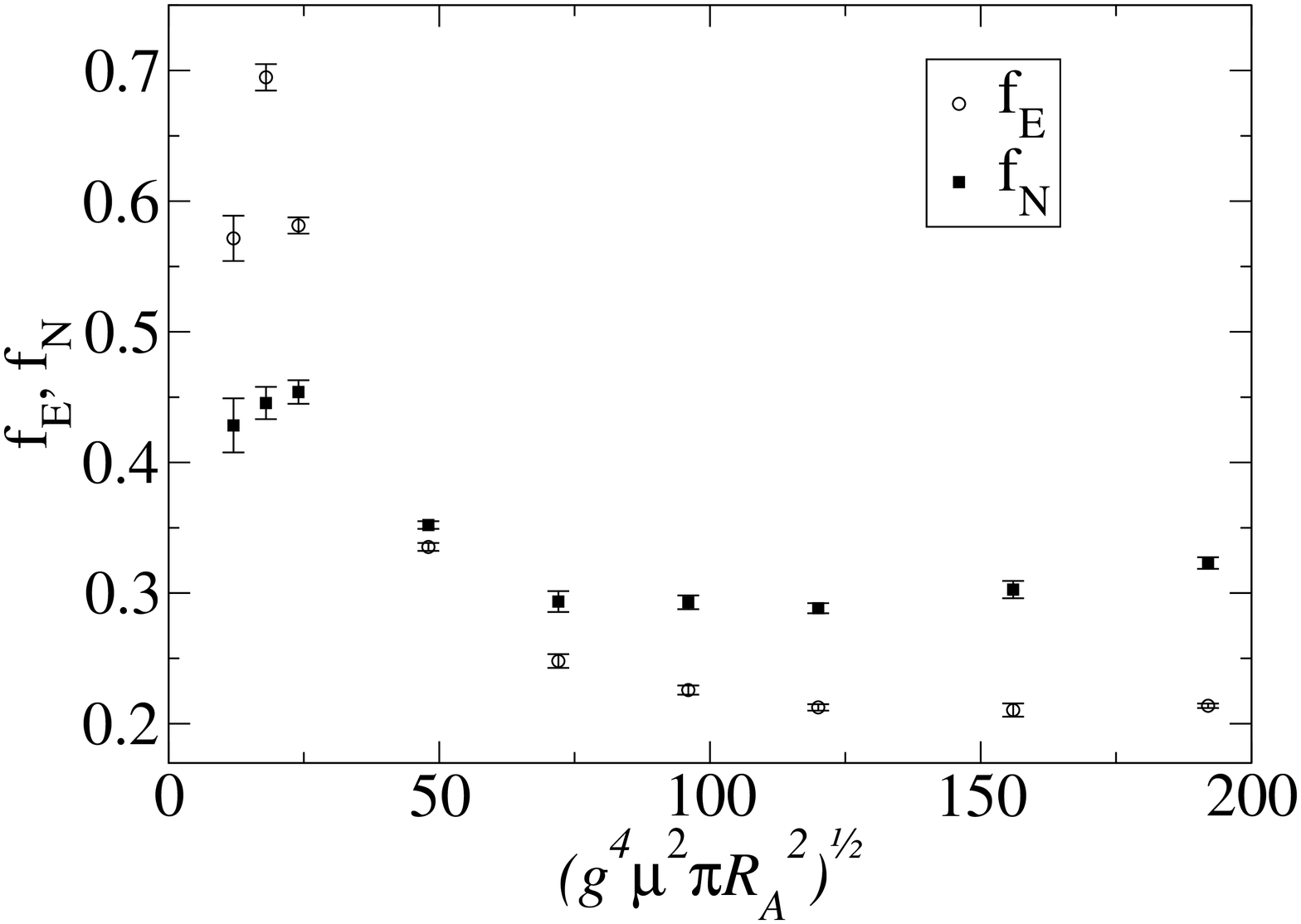}
\includegraphics[width=0.49\textwidth]{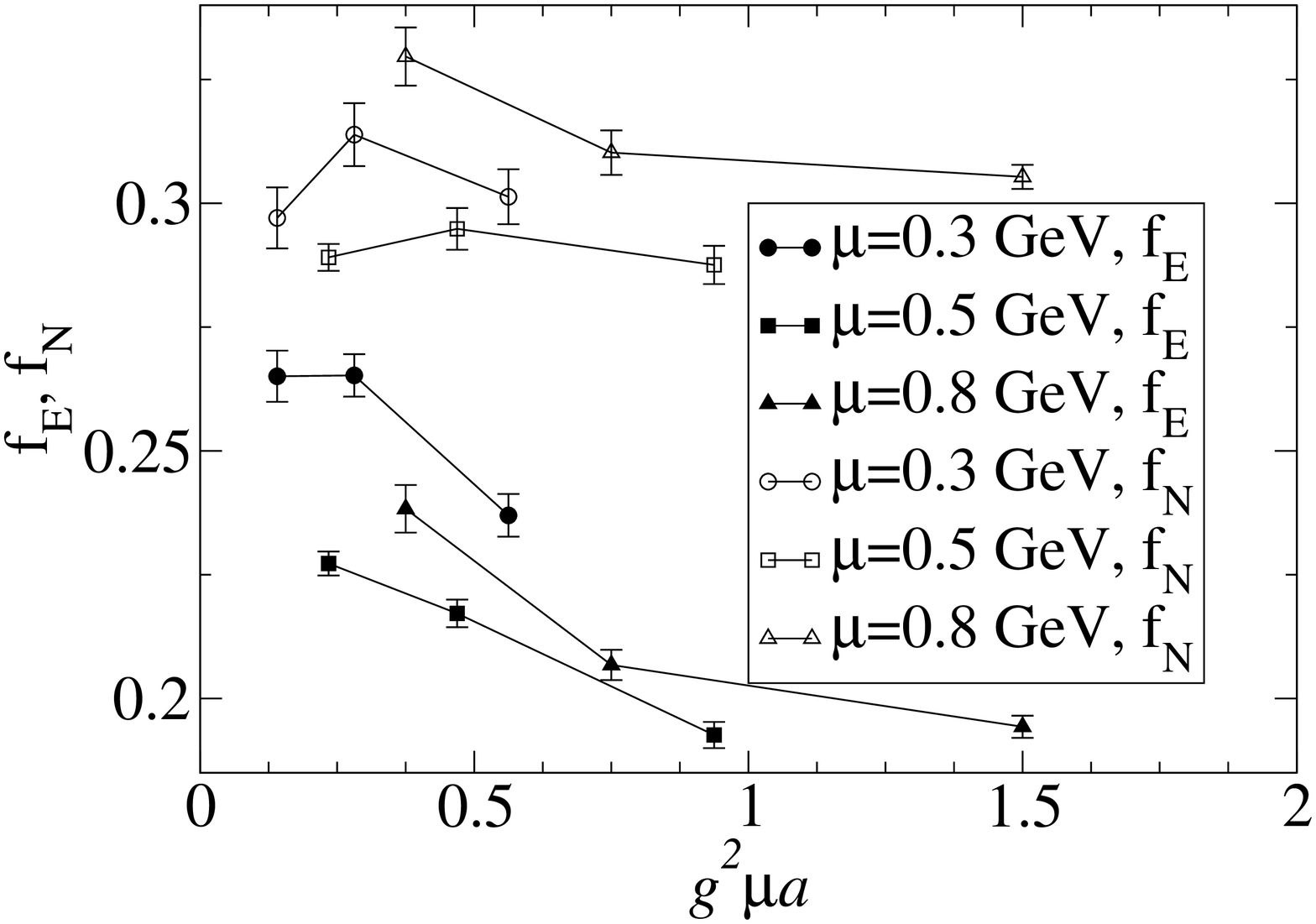}
\end{center}
\caption{Left: The dependence of $f_E$ and $f_N$ on the field stregth
paramater $g^2 \mu \ra$. Right: The dependence of $f_E$ and $f_N$
on lattice spacing for different values of $\mu$.}\label{fig:saturcontlim}
\end{figure}

\begin{figure}
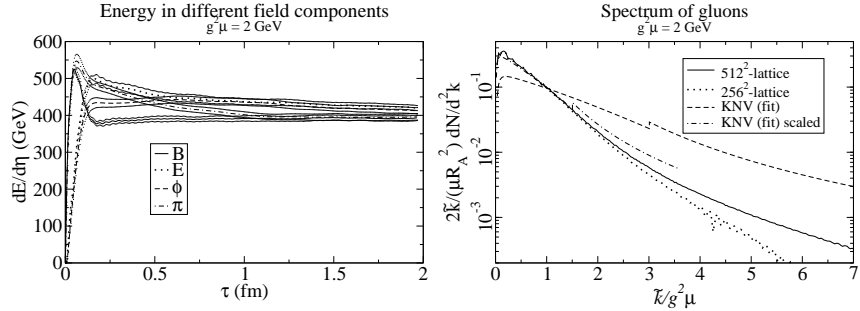

\begin{center}
\noindent
\includegraphics[width=0.49\textwidth]{lappi_fig4.eps}
\includegraphics[width=0.49\textwidth]{lappi_fig5.eps}
\end{center}
\caption{Left: the energy per unit rapidity is evenly distributed between
different field components and almost constant after a very short time
of the order of $\frac{1}{g^2 \mu}.$ This means that the 3+1 dimensional
energy density decreases as $\varepsilon \sim \tau^{-1}$
Right: The differential multiplicity for two different transverse lattice
sizes. The curves labeled ``KNV'' are a fit to the numerical result of
Ref.~\protect\refcite{Krasnitz:2001qu}, and KNV ``scaled'' after these results have been
corrected (see Ref.~\protect\refcite{Krasnitz:2003jw}).
}\label{fig:allennormvert}
\end{figure}

Let us define the dimensionless ratios describing the energy and multiplicity:
\begin{equation}
f_E = \frac{\ud E / \ud \eta}{g^4 \mu^3 \pi \ra^2 }
\quad
f_N = \frac{\ud N / \ud \eta}{g^2 \mu^2 \pi \ra^2 }
\end{equation}
These dimensionless ratios depend, apart from the lattice spacing $a$,
only on one dimensionless parameter characterising the field
strength,  $g^2 \mu \ra$.
As can be seen from Fig.~\ref{fig:saturcontlim},
for strong enough fields ($g^2 \mu \ra \gtrsim 50$) both
$f_E$ and $f_N$ are approximately independent of $g^2 \mu \ra$ and the
lattice spacing. Figure~\ref{fig:allennormvert} shows the energy as a function
of time in different field components and the spectrum of the produced gluons.

\subsection{Phenomenology}

At the level of the present discussion $g^2 \mu$ is still a free parameter
that needs to be fixed from the experimental data. One can distinguish between
two broad scenarios for relating the measured results to the calculated
initial state quantities.

\textbf{Hydro scenario} If the system thermalises fast, one
can use ideal hydrodynamics to follow its subsequent evolution.
In this scenario entropy and thus multiplicity are approximately 
conserved, but the transverse energy decreases by a factor
of $\sim 3$ due to $p\ud V$-work going down the beampipe.
The corresponding value for the saturation scale is 
$g^2 \mu \approx 1.9 \gev$.

\textbf{Free streaming scenario} Assuming that the gluons interact very
weakly and do not thermalize, one can argue that longitudinal pressure is
negligible and thus the transverse energy is conserved. Assuming
energy conservation one gets a lower value for the
saturation scale: $g^2 \mu \sim 1.4 \gev$. This lower value means that
the multiplicity must increase during thermalisation or hadronisation
of the system approximately by a factor of 2.

At least in the first case a prediction for the multiplicity at central
rapidities at the LHC is straightforward; one expects to see
$\left(\frac{5500 \gev}{130\gev} \right)^{0.3} \times 1000 \approx 3000$ 
particles per unit rapidity\footnote{Total, including neutral particles. 
At this level of approximation,
$N_\textrm{ch}\approx \frac{2}{3} N_\textrm{tot}$.}.

\section{Quark pair production}

Given the classical fields corresponding to gluon production
a natural question to ask is: how are quark-antiquark pairs produced by
these color fields? Formally quark production is suppressed by
$\alpha_\textrm{s}$ and group theory factors compared to gluons, so 
in a first approximation we should be able to treat the quarks as a small
perturbation and neglect their backreaction on the color fields.
Heavy quark production  is calculable already in perturbation
theory\footnote{In the weak field limit quark pair production from
this classical field model reduces to a known result in $k_\mathrm{T}$-factorised 
perturbation theory\cite{Gelis:2003vh}.},
but one can ask how much the strong color fields change the result.
Understanding light quark production would address the question of
chemical equilibration; turning the color glass condensate into a
\emph{quark} gluon plasma.

The calculation of quark pair production, outlined in more detail
in Ref.~\refcite{qqbartemp}, proceeds by solving the Dirac equation in the
background color field of the two nuclei. This can be done analytically
for QED\footnote{The QED calculation, of interest for
ultraperipheral collisions, is done
e.g. by Baltz and McLerran\cite{Baltz:1998zb} and others.
Baltz, Gelis, McLerran and Peshier\cite{Baltz:2001dp}
discuss the theory in more detail.}.
The initial condition for $t \to -\infty$ is a negative energy plane wave.
Similarly to the QED case, one can find analytically the solution for
the regions $x^\pm>0, \ x^\mp<0$. These then give the initial condition
at $\tau=0$ for a numerical solution of the Dirac equation in the 
forward light cone $\tau>0$. To find the number of quark pairs 
one then projects the numerically calculated wave function to a positive
energy plane wave at some sufficiently large time $\tau$.

A major technical challenge in this calculation is the coordinate system.
In order to include the hard sources of the color fields, the colliding nuclei, 
only in the initial condition of the numerical calculation, one wants
to use the proper time $\tau$ instead of the Minkowski time $t$. Unlike
the gluon production case, where one was able to assume strict boost invariance,
one now has a nontrivial correlation between the rapidities of the quark and
the antiquark. Thus, although the background gauge field is boost invariant,
one must solve the Dirac equation 3+1 dimensions.

\subsection{1+1-d toy model}

The longitudinal direction is the one posing the most technical problems.
To understand how to handle the longitudinal dimension
we can construct a 1+1-dimensional toy model without the transverse dimensions.
In 1+1-dimensions Dirac matrices are 2-dimensional. In the temporal
gauge $A_\tau=0$ that we have been using throughout the calculation there
is only one component, $A_\eta,$ in the external gauge field. The mass
$m^2_\textrm{eff}$ in the 1+1-dimensional model corresponds to the
transverse mass $\kk^2 + m^2$ of the full theory.

The initial condition at  $\tau = 0$ involves a
longitudinal momentum scale (the longitudinal momentum of the incoming
antiquark) and thus we must use a dimensionful longitudinal variable to
be able to represent the initial condition. We choose to take
as our coordinates $\tau$ and $z$. The Dirac equation becomes
\begin{equation}\label{eq:eomtauz}
\dtau \psi = \frac{\sqrt{\tau^2+z^2}+ \gt \gz z}{\tau}
\left(-\gt \gz \partial_z \psi 
- im_\textrm{eff} \gt \psi \right) - i \gt \gz   g \frac{A_\eta}{\tau} \psi.
\end{equation}
Because of the way the coefficients depend explicitly on the coordinates 
the discretisation of this equation is potentially very unstable and
we have to use an explicit discretisation method.

Let us take the background field as\footnote{
This is the correct time dependence of the 
perturbative solution to the gauge field eom's\cite{Kovner:1995ts}.}:
\begin{equation}
A_\eta = c \qs \tau J_1 (\qs \tau).
\end{equation}

\begin{figure}
\begin{center}
\includegraphics[width= 0.7\textwidth]{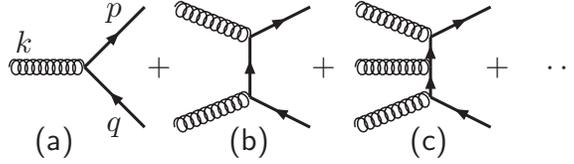}
\end{center}
\caption{Diagrams contributing to the quark pair prodction amplitude in the
1+1-dimensional toy model.}\label{fig:diags}
\end{figure}

\begin{figure}
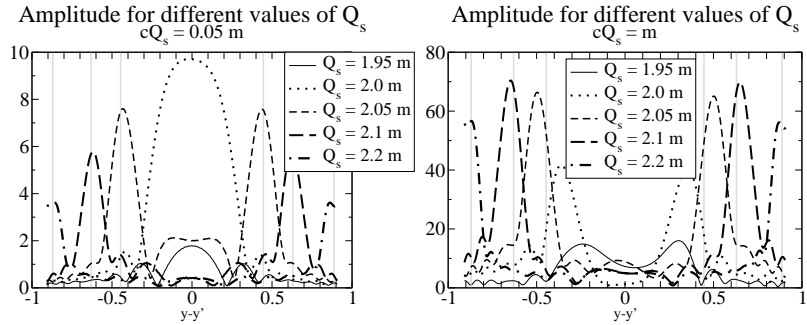

\begin{center}
\noindent
\includegraphics[height=120pt]{lappi_fig7.eps}
\includegraphics[height=120pt]{lappi_fig8.eps}
\end{center}
\caption{Absulute value of the quark pair production amplitude for different
values of the oscillation scale $\qs$. Left: weak fields, the peaks are
at the location explained by Eq.~\nr{eq:resonans}. Right: strong fields.
}\label{fig:ampli}
\end{figure}

For weak fields $c \ll 1$ we can compute the amplitude for quark pair production 
using the first order in perturbation theory (diagram (a) in Fig.~\ref{fig:diags}).
The result is a peak at
\begin{equation}\label{eq:resonans}
2k^+ k^- = (p+q)^2 = 2m_\textrm{eff}^2 \left(1 + \cosh (\Delta y) \right) =  \qs^2.
\end{equation}
As can be seen from Fig.~\ref{fig:ampli}, for weak fields our numerical computation
reproduces this perturbative peak. For stronger fields the numerical 
solution sums over all the diagrams in Fig.~\ref{fig:diags} and  the position 
of the peak is shifted.
In the full 3+1-dimensional case this peak is washed out by integration
over the relative transverse momentum of the pair.

A rapid back-of-the envelope estimate suggests that with transverse lattices
of the order of $256^2$ points (lattices from $128^2$ to $256^2$ were used
in the gluon production computation) one could just manage to have a large enough
lattice in the z-direction with the computers available to us. Then the computation
should in principle be repeated for all the different values of  $\pp$ (of the antiquark). 
As this would be prohibitively expensive in terms of CPU-time one will,
however, probably have to do with an interpolation from a reasonable amount 
of points on the $\pp$ lattice.

\section{Conclusions}

Production of gluons has been computed numerically in the classical field
model for heavy ion collisions, and the results are in rough agreement
with RHIC observations, although not very conclusive due to the uncertainty
in the numerical value of the saturation scale.
A numerical calculation of the number of quark pairs produced from these classical
fields is under way. It is hoped that this computation will tell us something about the
how a purely gluonic system can transform into a plasma of both quarks and gluons.
Understanding kinetic thermalisation in this context will require treatment of the
longitudinal dimension and implementing initial conditions from the JIMWLK equation.

\section*{Acknowledgments}
The author wishes to thank K. Kajantie and F. Gelis for collaboration in the work
presented here and  K. Rummukainen, K. Tuominen, A. Krasnitz, Y. Nara and R. Venugopalan
for discussions. This work was supported by the Finnish Cultural Foundation and the
Magnus Ehrnrooth Foundation.


\begin{thebibliography}{10}

\bibitem{kolb}
P.~Kolb,
\newblock These proceedings, nucl-th/0407066.

\bibitem{Baier:2000sb}
R.~Baier, A.~H. Mueller, D.~Schiff and D.~T. Son,
\newblock Phys. Lett. {\bf B502}, 51 (2001), [hep-ph/0009237].

\bibitem{arnold}
P.~Arnold,
\newblock These proceedings, hep-ph/0409002.

\bibitem{Golec-Biernat:1998js}
K.~Golec-Biernat and M.~Wusthoff,
\newblock Phys. Rev. {\bf D59}, 014017 (1999), [hep-ph/9807513].

\bibitem{McLerran:1994ni}
L.~D. McLerran and R.~Venugopalan,
\newblock Phys. Rev. {\bf D49}, 2233 (1994), [hep-ph/9309289].

\bibitem{weigert}
H.~Weigert,
\newblock These proceedings.

\bibitem{Eskola:1999fc}
K.~J. Eskola, K.~Kajantie, P.~V. Ruuskanen and K.~Tuominen,
\newblock Nucl. Phys. {\bf B570}, 379 (2000), [hep-ph/9909456].

\bibitem{Krasnitz:2002mn}
A.~Krasnitz, Y.~Nara and R.~Venugopalan,
\newblock Nucl. Phys. {\bf A717}, 268 (2003), [hep-ph/0209269].

\bibitem{Kovchegov:2000hz}
Y.~V. Kovchegov,
\newblock Nucl. Phys. {\bf A692}, 557 (2001), [hep-ph/0011252].

\bibitem{Rummukainen:2003ns}
K.~Rummukainen and H.~Weigert,
\newblock Nucl. Phys. {\bf A739}, 183 (2004), [hep-ph/0309306].

\bibitem{Lam:2002mg}
C.~S. Lam, G.~Mahlon and W.~Zhu,
\newblock Phys. Rev. {\bf D66}, 074005 (2002), [hep-ph/0207058].

\bibitem{Kovner:1995ts}
A.~Kovner, L.~D. McLerran and H.~Weigert,
\newblock Phys. Rev. {\bf D52}, 3809 (1995), [hep-ph/9505320].

\bibitem{Krasnitz:1998ns}
A.~Krasnitz and R.~Venugopalan,
\newblock Nucl. Phys. {\bf B557}, 237 (1999), [hep-ph/9809433].

\bibitem{Krasnitz:2001qu}
A.~Krasnitz, Y.~Nara and R.~Venugopalan,
\newblock Phys. Rev. Lett. {\bf 87}, 192302 (2001), [hep-ph/0108092].

\bibitem{Lappi:2003bi}
T.~Lappi,
\newblock Phys. Rev. {\bf C67}, 054903 (2003), [hep-ph/0303076].

\bibitem{Krasnitz:2003jw}
A.~Krasnitz, Y.~Nara and R.~Venugopalan,
\newblock Nucl. Phys. {\bf A727}, 427 (2003), [hep-ph/0305112].

\bibitem{Gelis:2003vh}
F.~Gelis and R.~Venugopalan,
\newblock Phys. Rev. {\bf D69}, 014019 (2004), [hep-ph/0310090].

\bibitem{qqbartemp}
F.~Gelis, K.~Kajantie and T.~Lappi,
\newblock hep-ph/0409058.

\bibitem{Baltz:1998zb}
A.~J. Baltz and L.~D. McLerran,
\newblock Phys. Rev. {\bf C58}, 1679 (1998), [nucl-th/9804042].

\bibitem{Baltz:2001dp}
A.~J. Baltz, F.~Gelis, L.~D. McLerran and A.~Peshier,
\newblock Nucl. Phys. {\bf A695}, 395 (2001), [nucl-th/0101024].

\end{thebibliography}
\end{document}